\documentclass[aps,prl,twocolumn,showpacs,superscriptaddress,letterpaper,amsmath,amssymb]{revtex4}  

\usepackage{graphicx}
\usepackage{dcolumn}
\usepackage{bm}

\RequirePackage{xspace}
\usepackage{relsize}

\def\Kbar  {\kern 0.2em\overline{\kern -0.2em K}{}\xspace}

\def\Bbar    {\kern 0.18em\overline{\kern -0.18em B}{}\xspace}

\def\bprime  {\ensuremath{b^{\prime}}\xspace}

\def\Qbar    {\kern 0.08em\overline{\kern -0.08em Q}{}\xspace}

\def\invfb   {\ensuremath{\mbox{\,fb}^{-1}}\xspace}
\newcommand{\mev}{\ensuremath{\mathrm{\,Me\kern -0.1em V}}\xspace}
\newcommand{\mevc}{\ensuremath{{\mathrm{\,Me\kern -0.1em V\!/}c}}\xspace}
\newcommand{\mevcc}{\ensuremath{{\mathrm{\,Me\kern -0.1em V\!/}c^2}}\xspace}
\newcommand{\gev}{\ensuremath{\mathrm{\,Ge\kern -0.1em V}}\xspace}
\newcommand{\gevc}{\ensuremath{{\mathrm{\,Ge\kern -0.1em V\!/}c}}\xspace}
\newcommand{\gevcnospace}{\ensuremath{{\mathrm{\,Ge\kern -0.1em V\!/}c}}}
\newcommand{\gevcc}{\ensuremath{{\mathrm{\,Ge\kern -0.1em V\!/}c^2}}\xspace}

\def\missET {\mbox{\ensuremath{\not\!\! E_T}}}

\begin{document}
\bibliographystyle{apsrev}

\title{{Search for heavy bottom-like quarks \\decaying to an electron
    or muon and jets in $p\bar{p}$ collisions at $\sqrt{s}=1.96$ TeV}}

\affiliation{Institute of Physics, Academia Sinica, Taipei, Taiwan 11529, Republic of China} 
\affiliation{Argonne National Laboratory, Argonne, Illinois 60439, USA} 
\affiliation{University of Athens, 157 71 Athens, Greece} 
\affiliation{Institut de Fisica d'Altes Energies, ICREA, Universitat Autonoma de Barcelona, E-08193, Bellaterra (Barcelona), Spain} 
\affiliation{Baylor University, Waco, Texas 76798, USA} 
\affiliation{Istituto Nazionale di Fisica Nucleare Bologna, $^z$University of Bologna, I-40127 Bologna, Italy} 
\affiliation{University of California, Davis, Davis, California 95616,
USA} 
\affiliation{University of California, Irvine, Irvine, California 92697, USA}
\affiliation{University of California, Los Angeles, Los Angeles, California 90024, USA} 
\affiliation{Instituto de Fisica de Cantabria, CSIC-University of Cantabria, 39005 Santander, Spain} 
\affiliation{Carnegie Mellon University, Pittsburgh, Pennsylvania 15213, USA} 
\affiliation{Enrico Fermi Institute, University of Chicago, Chicago, Illinois 60637, USA}
\affiliation{Comenius University, 842 48 Bratislava, Slovakia; Institute of Experimental Physics, 040 01 Kosice, Slovakia} 
\affiliation{Joint Institute for Nuclear Research, RU-141980 Dubna, Russia} 
\affiliation{Duke University, Durham, North Carolina 27708, USA} 
\affiliation{Fermi National Accelerator Laboratory, Batavia, Illinois 60510, USA} 
\affiliation{University of Florida, Gainesville, Florida 32611, USA} 
\affiliation{Laboratori Nazionali di Frascati, Istituto Nazionale di Fisica Nucleare, I-00044 Frascati, Italy} 
\affiliation{University of Geneva, CH-1211 Geneva 4, Switzerland} 
\affiliation{Glasgow University, Glasgow G12 8QQ, United Kingdom} 
\affiliation{Harvard University, Cambridge, Massachusetts 02138, USA} 
\affiliation{Division of High Energy Physics, Department of Physics, University of Helsinki and Helsinki Institute of Physics, FIN-00014, Helsinki, Finland} 
\affiliation{University of Illinois, Urbana, Illinois 61801, USA} 
\affiliation{The Johns Hopkins University, Baltimore, Maryland 21218, USA} 
\affiliation{Institut f\"{u}r Experimentelle Kernphysik, Karlsruhe Institute of Technology, D-76131 Karlsruhe, Germany} 
\affiliation{Center for High Energy Physics: Kyungpook National University, Daegu 702-701, Korea; Seoul National University, Seoul 151-742, Korea; Sungkyunkwan University, Suwon 440-746, Korea; Korea Institute of Science and Technology Information, Daejeon 305-806, Korea; Chonnam National University, Gwangju 500-757, Korea; Chonbuk National University, Jeonju 561-756, Korea} 
\affiliation{Ernest Orlando Lawrence Berkeley National Laboratory, Berkeley, California 94720, USA} 
\affiliation{University of Liverpool, Liverpool L69 7ZE, United Kingdom} 
\affiliation{University College London, London WC1E 6BT, United Kingdom} 
\affiliation{Centro de Investigaciones Energeticas Medioambientales y Tecnologicas, E-28040 Madrid, Spain} 
\affiliation{Massachusetts Institute of Technology, Cambridge, Massachusetts 02139, USA} 
\affiliation{Institute of Particle Physics: McGill University, Montr\'{e}al, Qu\'{e}bec, Canada H3A~2T8; Simon Fraser University, Burnaby, British Columbia, Canada V5A~1S6; University of Toronto, Toronto, Ontario, Canada M5S~1A7; and TRIUMF, Vancouver, British Columbia, Canada V6T~2A3} 
\affiliation{University of Michigan, Ann Arbor, Michigan 48109, USA} 
\affiliation{Michigan State University, East Lansing, Michigan 48824, USA}
\affiliation{Institution for Theoretical and Experimental Physics, ITEP, Moscow 117259, Russia}
\affiliation{University of New Mexico, Albuquerque, New Mexico 87131,
USA} 
\affiliation{Northwestern University, Evanston, Illinois 60208, USA} 
\affiliation{The Ohio State University, Columbus, Ohio 43210, USA} 
\affiliation{Okayama University, Okayama 700-8530, Japan} 
\affiliation{Osaka City University, Osaka 588, Japan} 
\affiliation{University of Oxford, Oxford OX1 3RH, United Kingdom} 
\affiliation{Istituto Nazionale di Fisica Nucleare, Sezione di Padova-Trento, $^{aa}$University of Padova, I-35131 Padova, Italy} 
\affiliation{LPNHE, Universite Pierre et Marie Curie/IN2P3-CNRS, UMR7585, Paris, F-75252 France} 
\affiliation{University of Pennsylvania, Philadelphia, Pennsylvania 19104, USA}
\affiliation{Istituto Nazionale di Fisica Nucleare Pisa, $^{bb}$University of Pisa, $^{cc}$University of Siena and $^{dd}$Scuola Normale Superiore, I-56127 Pisa, Italy} 
\affiliation{University of Pittsburgh, Pittsburgh, Pennsylvania 15260, USA} 
\affiliation{Purdue University, West Lafayette, Indiana 47907, USA} 
\affiliation{University of Rochester, Rochester, New York 14627, USA} 
\affiliation{The Rockefeller University, New York, New York 10065, USA} 
\affiliation{Istituto Nazionale di Fisica Nucleare, Sezione di Roma 1, $^{ee}$Sapienza Universit\`{a} di Roma, I-00185 Roma, Italy} 

\affiliation{Rutgers University, Piscataway, New Jersey 08855, USA} 
\affiliation{Texas A\&M University, College Station, Texas 77843, USA} 
\affiliation{Istituto Nazionale di Fisica Nucleare Trieste/Udine, I-34100 Trieste, $^{ff}$University of Trieste/Udine, I-33100 Udine, Italy} 
\affiliation{University of Tsukuba, Tsukuba, Ibaraki 305, Japan} 
\affiliation{Tufts University, Medford, Massachusetts 02155, USA} 
\affiliation{University of Virginia, Charlottesville, VA  22906, USA}
\affiliation{Waseda University, Tokyo 169, Japan} 
\affiliation{Wayne State University, Detroit, Michigan 48201, USA} 
\affiliation{University of Wisconsin, Madison, Wisconsin 53706, USA} 
\affiliation{Yale University, New Haven, Connecticut 06520, USA} 
\author{T.~Aaltonen}
\affiliation{Division of High Energy Physics, Department of Physics, University of Helsinki and Helsinki Institute of Physics, FIN-00014, Helsinki, Finland}
\author{B.~\'{A}lvarez~Gonz\'{a}lez$^v$}
\affiliation{Instituto de Fisica de Cantabria, CSIC-University of Cantabria, 39005 Santander, Spain}
\author{S.~Amerio}
\affiliation{Istituto Nazionale di Fisica Nucleare, Sezione di Padova-Trento, $^{aa}$University of Padova, I-35131 Padova, Italy} 

\author{D.~Amidei}
\affiliation{University of Michigan, Ann Arbor, Michigan 48109, USA}
\author{A.~Anastassov}
\affiliation{Northwestern University, Evanston, Illinois 60208, USA}
\author{A.~Annovi}
\affiliation{Laboratori Nazionali di Frascati, Istituto Nazionale di Fisica Nucleare, I-00044 Frascati, Italy}
\author{J.~Antos}
\affiliation{Comenius University, 842 48 Bratislava, Slovakia; Institute of Experimental Physics, 040 01 Kosice, Slovakia}
\author{G.~Apollinari}
\affiliation{Fermi National Accelerator Laboratory, Batavia, Illinois 60510, USA}
\author{J.A.~Appel}
\affiliation{Fermi National Accelerator Laboratory, Batavia, Illinois 60510, USA}
\author{A.~Apresyan}
\affiliation{Purdue University, West Lafayette, Indiana 47907, USA}
% ...
\collaboration{CDF Collaboration\footnote{With visitors from $^a$University of Massachusetts Amherst, Amherst, Massachusetts 01003,
$^b$Istituto Nazionale di Fisica Nucleare, Sezione di Cagliari, 09042 Monserrato (Cagliari), Italy,
$^c$University of California Irvine, Irvine, CA  92697, 
$^d$University of California Santa Barbara, Santa Barbara, CA 93106
$^e$University of California Santa Cruz, Santa Cruz, CA  95064,
$^f$CERN,CH-1211 Geneva, Switzerland,
$^g$Cornell University, Ithaca, NY  14853, 
$^h$University of Cyprus, Nicosia CY-1678, Cyprus, 
$^i$University College Dublin, Dublin 4, Ireland,
$^j$University of Fukui, Fukui City, Fukui Prefecture, Japan 910-0017,
$^k$Universidad Iberoamericana, Mexico D.F., Mexico,
$^l$Iowa State University, Ames, IA  50011,
$^m$University of Iowa, Iowa City, IA  52242,
$^n$Kinki University, Higashi-Osaka City, Japan 577-8502,
$^o$Kansas State University, Manhattan, KS 66506,
$^p$University of Manchester, Manchester M13 9PL, England,
$^q$Queen Mary, University of London, London, E1 4NS, England,
$^r$Muons, Inc., Batavia, IL 60510,
$^s$Nagasaki Institute of Applied Science, Nagasaki, Japan, 
$^t$National Research Nuclear University, Moscow, Russia,
$^u$University of Notre Dame, Notre Dame, IN 46556,
$^v$Universidad de Oviedo, E-33007 Oviedo, Spain, 
$^w$Texas Tech University, Lubbock, TX  79609, 
$^x$Universidad Tecnica Federico Santa Maria, 110v Valparaiso, Chile,
$^y$Yarmouk University, Irbid 211-63, Jordan,
$^{gg}$On leave from J.~Stefan Institute, Ljubljana, Slovenia, 
}}
\noaffiliation

\begin{abstract}
We report the most sensitive direct search for pair production of
fourth-generation bottom-like chiral quarks ($b'$) each decaying promptly to
$tW$. We search for an excess of events with an electron or
muon, at least five jets (one indentified as due to a $b$ or $c$ quark) and an imbalance of transverse momentum using data from $p\bar{p}$ collisions collected
by the CDF II detector at Fermilab with an integrated luminosity of
4.8 fb$^{-1}$. We observe events consistent with background
expectation and calculate upper limits on the $b'$ pair production
cross section ($\sigma_{b\bar{b'}}\lesssim 30$ fb for $m_{b'}>$375 GeV/$c^2$) and exclude
$m_{b'}<372$ \gevcc  at 95\% confidence level.
\end{abstract}



\maketitle

\date{\today}

The standard model (SM) of particle physics accomodates three generations
of fundamental fermions, but is agnostic on the issue of a fourth
generation. Precision  measurements in the electroweak sector are not inconsistent with a
fourth-generation of fermions if there is a 50-100 GeV$/c^2$ splitting in the
quark and lepton masses~\cite{4fmasses}.  A four-generation model
\cite{Frampton:1999xi} could provide a source of particle-antiparticle
asymmetry large enough to account for the baryon asymmetry of the
universe~\cite{Hou:2009BAU}, and accomodate a heavier Higgs boson
(the source of mass generation) than a three-generation
model~\cite{Kribs:2007nz}.  Direct searches for production of chiral fourth
generation quarks restrict their masses to be greater than 335
GeV$/c^2$~\cite{cdftp} for an up-type quark $t'$ decaying via $t'\rightarrow
Wq$ and 338 GeV$/c^2$~\cite{cdfbp} for a down-type quark $b'$ decaying via
$b'\rightarrow tW$.

This Letter reports a search for pair-production via strong
interactions of a heavy
chiral~\cite{chiral} bottom-like quark, $b'$,
 followed by prompt decay to a $t$
quark and a $W$ boson with $\mathcal{B}(b'\rightarrow Wt)=100$\%. The assumption that $b'$ decays exclusively to
$t W$ is reasonable if
the coupling to light quarks is small, as expected from precision
meson-mixing measurements~\cite{mesonmixing}, and in the hypothesis that $m_{\bprime} >$
$m_{t}+m_{W}$.  In the case that the branching fraction deviates
from 100\%, the limits can be interpreted under
different assumptions~\cite{masspaper}. Previous searches considered the mode in which
two same-charge $W$ bosons decayed leptonically~\cite{cdfbp},
which gives a low-background signature but a low selection
efficiency due to the small $W\rightarrow \ell\nu$ branching ratio.
We consider the mode $ b'\bar{b'} \rightarrow WtW\bar{t} \rightarrow
WWb WW\bar{b} \rightarrow \ell\nu qq' b qq' qq' b$ in
which one $W$ boson decays leptonically (including $\tau$ decays to
$e$ or $\mu$) and the remaining three $W$ bosons decay hadronically,
giving a selection efficiency nearly four times the previous search.
The larger SM backgrounds can be separated from a
potential signal by comparing the total reconstructed transverse momentum in
the event.


Events were recorded by CDF II~\cite{cdf,coordinates}, a general purpose detector
designed to study collisions at the Fermilab Tevatron
$p\overline{p}$ collider at $\sqrt{s}=1.96$ TeV. A charged-particle tracking system immersed
in a 1.4~T magnetic field consists of a silicon microstrip tracker and
a drift chamber. Electromagnetic and hadronic calorimeters surround
the tracking system and measure particle energies. Drift chambers
located outside the calorimeters detect muons.  We use a data sample
corresponding to an integrated luminosity of 4.8$\pm 0.3$\invfb.

The data acquisition system is triggered by $e$ or $\mu$ candidates~\cite{leptonid}
with transverse momentum ($p_T$\cite{coordinates})  greater than $
18$\gevc.  Electrons and muons are reconstructed offline and selected if they
have a pseudorapidity ($\eta$\cite{coordinates}) magnitude less than
1.1, $p_T \ge 20$ GeV$/c$ and satisfy the standard CDF identification and
isolation requirements~\cite{leptonid}.  Jets are reconstructed in the
calorimeter using the {\sc jetclu}~\cite{jetclu}  algorithm with a
clustering radius of 0.4 in azimuth-pseudorapidity space and corrected using the standard techniques~\cite{jes}. Jets are
selected if they have $p_T \ge 15$ GeV$/c$ and $|\eta| < 2.4$. Each jet
is considered for heavy-flavor tagging using
the default CDF $b$-jet identification algorithm ({\sc secvtx}\cite{secvtx})
that searches in the jet for a
secondary vertex which results from the displaced decay of a
$B$-hadron inside the jet. Missing transverse momentum~\cite{metdef}
is reconstructed using fully corrected
calorimeter and muon information~\cite{leptonid}.

Production and decay of $b'$ pairs would appear as events with a
charged lepton and missing transverse momentum from one leptonically decaying $W$, and a large
number of jets from the two $b$ quarks and the hadronic decays of the
other three $W$ bosons.   We select events with exactly one electron or
muon, at least five jets, and at least 20 GeV/$c$ of
missing transverse momentum. At least one of the jets must be identified
as due to $b$ quark decay.  We find 357 events satisfying these requirements.

We model the production and decay of $b'$ pairs with {\sc
  madgraph}~\cite{madgraph}. Additional radiation, hadronization and
showering are described by {\sc pythia}~\cite{pythia}. The detector
response for all simulated samples is modeled by {\sc cdfsim}~\cite{cdfsim}. The signal
efficiency for the above requirements is approximately 10\%,
rising with $b'$ mass. There are eight quarks produced in the decay,
but the most likely number of reconstructed jets is six, as quarks
that are close together are likely to be merged into a single jet, and
some of the quarks produce jets which fall below the transverse
momentum threshold. Complete mass reconstruction
is therefore not possible in the majority of the events; instead, we
examine the event $H_T$, the scalar sum of the transverse momentum of
the lepton, jets and missing transverse momentum. This is
 well correlated with the mass of the heavy quark and serves as an
 approximate mass reconstruction.    

The dominant background (80\%) is top-quark pair production with
additional jets from initial or final state radiation. This background
can be distinguished from the signal as it has smaller $H_{T}$.  We model this background using
{\sc madgraph} $t\bar{t}$ production with $m_t = 172.5$ GeV/$c^2$ in
which radiation of up to three additional hard partons
(including heavy flavor)
are described explicitly using matrix-elements, and additional
radiation is described by the parton-shower; the {\sc mlm}~\cite{mlm} scheme is
used to match the matrix-element and parton-shower contributions.
This gives a precise description of events with $\le 7$ jets, where a $b'$ signal
would be expected. Events with eight jets and above are
described by the parton shower, which has significantly larger
systematic uncertainties. We normalize the $t\bar{t}$ background to the NLO cross section~\cite{nloa}, and
confirm that it is well modeled by examining $t\bar{t}$-dominated regions in the data.

 The second dominant background process ($\approx10$\%) is the associated production 
of $W$ boson and jets. Samples of simulated $W$+jets events with light- and heavy-flavor 
jets are generated using the {\sc alpgen}~\cite{alpgen} program, interfaced with parton-shower model 
from {\sc pythia}.  The $W+$jets samples are normalized to the measured $W$ cross section, 
with an additional multiplicative factor for the relative contribution of heavy- 
and light-flavor jets, the standard technique in measuring the
top-quark pair production cross section~\cite{secvtx}. Multi-jet
background ($\approx5$\%), in which a jet is
misreconstructed as a lepton, is modeled using a jet-triggered sample normalized
in a background-dominated region at low missing transverse momentum.
The remaining backgrounds, single top and diboson production, are modeled
using {\sc pythia} and normalized to next-to-leading order
cross sections~\cite{ewknlo}. The combined background expectation is 365$\pm
194$ events, including systematic and statistical uncertainties.

\begin{figure}[ht]
\begin{center}
\includegraphics[width=0.95\linewidth]{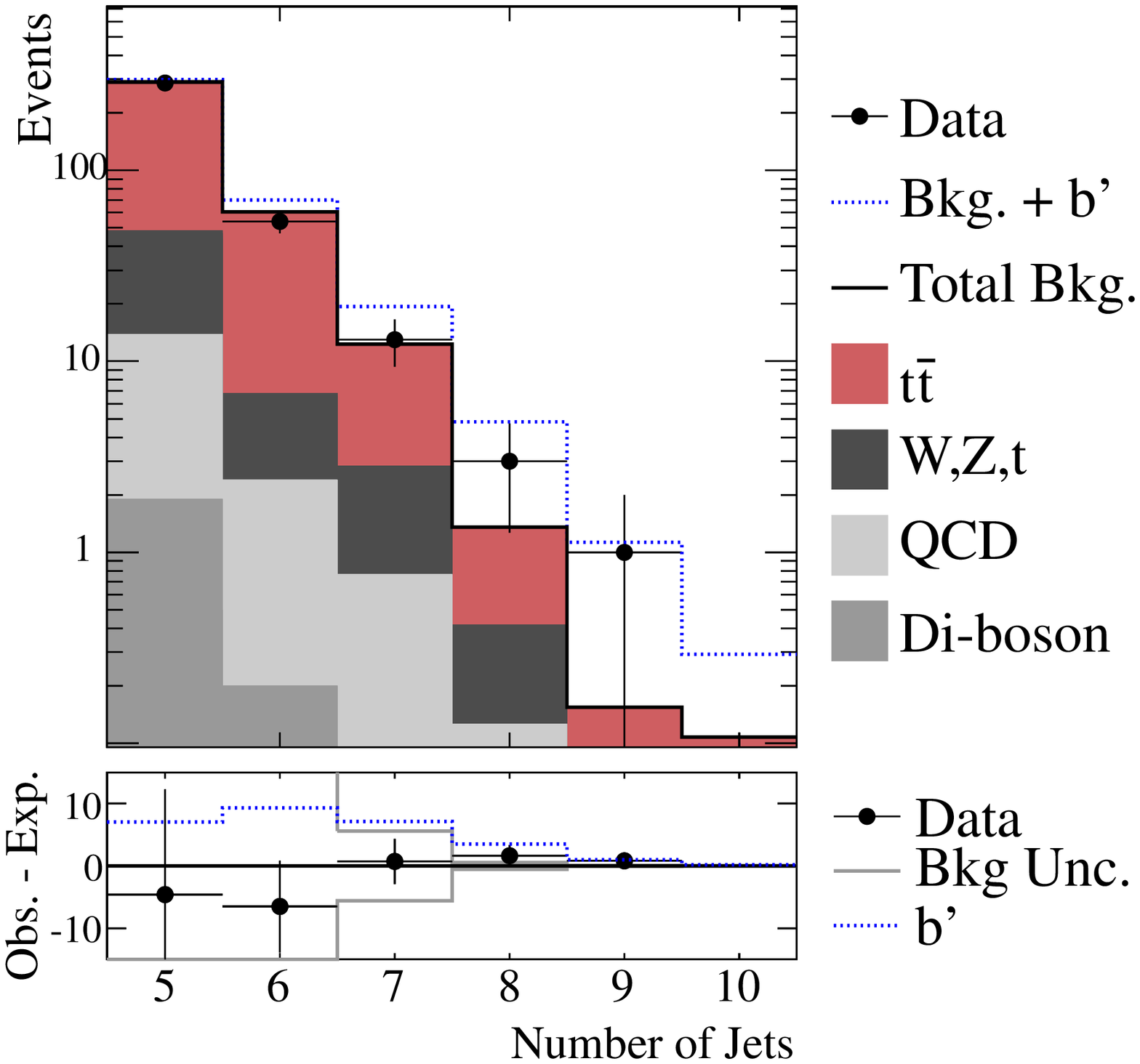}\\
\includegraphics[width=0.95\linewidth]{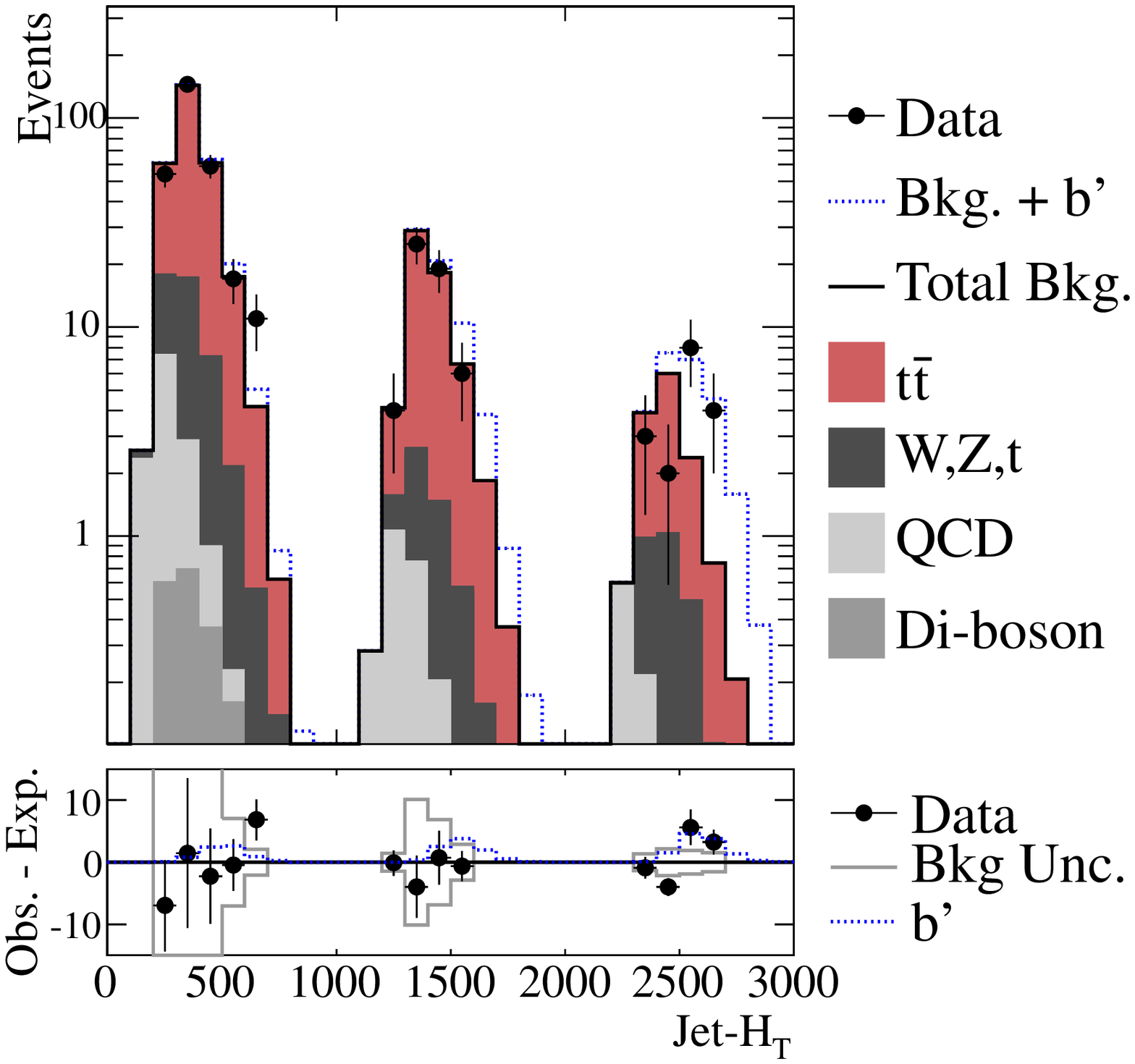}
\caption{ Distributions in jet multiplicity and Jet-$H_{T}$ (defined
  in the text). The example $b'$ signal has $m_{b'}=350$ GeV$/c^2$ and
  would have $29 \pm 4.5$ events expected in this sample.  Top pane is
  log scale; bottom pane shows the difference between expected and
  observed events
  on a linear scale, as well as the total uncertainty on the expected
  events.}
\label{fig:5j_njets}
\end{center}
\end{figure}

A $b'$ signal would be readily separated from the background both in the number of
jets and the $H_T$. To take advantage of both of
these characteristics, we introduce a variable ``Jet-$H_T$'',  which
equals $H_T$ for events with exactly 5 jets, Jet-$H_T = H_T+1000$ GeV
for events with exactly 6 jets,  and Jet-$H_T = H_T+2000$ GeV for
events with at least 7 jets. This is equivalent to a two-dimensional analysis in $N_{jets}$
and $H_T$.  Figure ~\ref{fig:5j_njets} shows the
distributions of an example $b'$ signal with $m_{b'}=350$ \gevcc and the backgrounds in Jet-$H_T$.

We consider several sources of systematic uncertainty on both the background
rates and distributions, as well as on the expectations for the
signal. Each affects the expected sensitivity to new physics expressed
as an expected cross section upper limit in the no-signal
assumption. The dominant systematic uncertainties are the jet energy scale~\cite{jes},
contributions from additional interactions, and descriptions of
initial and final state radiation~\cite{rad}.  In each case, we treat
the unknown underlying quantity as a nuisance parameter and measure the distortion of the
Jet-$H_T$ spectrum for positive and negative fluctuations. 
Each uncertainty weakens the
expected 95\% confidence level (C.L.) cross section upper limit by $\approx60$\%
individually. Additional uncertainty comes from parton distribution
functions (PDF) ~\cite{pdfs1,pdfs2}, the matching scale used
between the matrix-element and the parton shower, overall background
normalization, and uncertainties in performance of the $b$-quark
identification algorithm.  The overall impact on the expected
sensitivity is $\approx100$\% in the cross section and $\approx20$ GeV/$c^2$ on the expected mass limit.

 To validate the description of the backgrounds, we verify that the
low $H_{T}$ region is well-described where there is little signal
expected. See Table~\ref{tab:jetcont1ti}.   In events with $\ge 7$ jets, the observed $H_T$ is larger than predicted by
our background model. The number of observed events with  $\ge 7$ jets
and $H_T>500$ GeV
is 12 where we expect $3.4\pm 3.4$. However, the total number of
events observed in the low $H_T$ and high $H_T$ regions combined is
consistent with expectation.  Considering only the number of events in
the high $H_T$ regions and taking into account the systematic
uncertainties in the background prediction,  we see a more 
significant excess than that observed in the data in 12\% of simulated experiments.

\begin{table}
\caption{Expected and observed events in a background-dominated control
  region
  ($H_T < 400,450,500$ for $N_{jet}=5,6,\ge 7$, respectively) and in a
  signal-dominated region ($H_T > 400,450,500$ for $N_{jet}=5,6,\ge
  7$) for our selection (see text). Uncertainties are
  statistical and systematic, combined in quadrature.}
\label{tab:jetcont1ti}
\begin{center}
\begin{ruledtabular}
\begin{tabular}{llrlrlr}
 & \multicolumn{2}{c}{Control Region}  &
 \multicolumn{2}{c}{Signal Region}  &
 \multicolumn{2}{c}{Sum} \\
Jets & Exp. & Obs.  &  Exp.
 & Obs.  & Exp. & Obs. \\  \hline
\vspace{0.1cm} 5       & 207 $\pm\  125$ & 199   & 84  $\pm\  65$  & 87    & 291  $\pm\  190$        & 286\\

\vspace{0.1cm} 6       & 43 $\pm\  31$   & 40    & 18  $\pm\  12$   & 14    & 61  $\pm\  43$  & 54\\
$\ge 7$      & 11 $\pm\  3.9$     & 5     & 3.4  $\pm\  3.4$     & 12    & 14  $\pm\  7.1$    & 17\\
\end{tabular}
\end{ruledtabular}
\end{center}
\end{table}

\begin{figure}[ht]
\begin{center}
\includegraphics[width=0.95\linewidth]{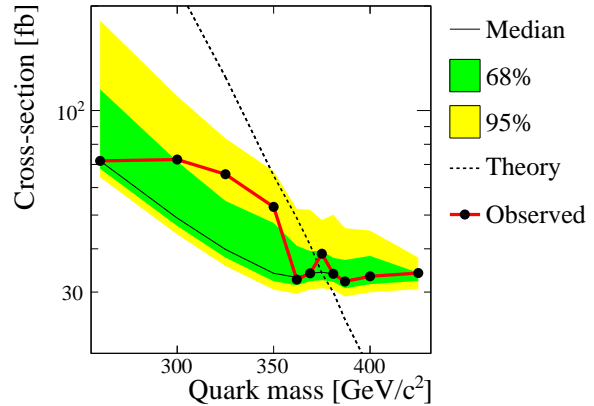}
\caption{ Upper limits on $b'$ production cross section at 95\% C.L.
  assuming $\mathcal{B}(b'\rightarrow Wt$)=100\%. Solid black
  line is the median expected upper limit in simulated experiments
  without $b'$ signal; green and yellow bands represent 68\% and 95\%
  of simulated experiments, respectively; solid red line is the observed
  limit.  Dashed black line is the NLO $b'$ production cross section~\cite{nloa,nlob}.}
\label{fig:limits}
\end{center}
\end{figure}

The full Jet-$H_T$ spectrum is used in the analysis. Since there is no evidence for the presence of $b'$ events in the
data, we calculate 95\% C.L. upper limits on the $b'$ production cross section,
by performing a binned maximum-likelihood fit in the Jet-$H_T$ variable,
allowing for systematic and statistical fluctuations via template
morphing~\cite{morph}. We  use the likelihood-ratio ordering
prescription~\cite{feldcous} to construct classical confidence
intervals in the theoretical cross section
by generating ensembles of simulated experiments that describe
expected fluctuations of statistical and systematic uncertainties on
both signal and backgrounds. The observed limits are consistent with expectation in
the background-only hypothesis and are given together with theoretical
next-to-leading-order (NLO) cross sections~\cite{nloa,nlob} in
Table~\ref{table:bprres} and shown in Fig.~\ref{fig:limits}. 

We convert upper limits on the pair-production cross sections to lower
limits on the fermion masses. The relative cross-section uncertainty of $\approx10$\% due to scale
and PDF uncertainties translates into  $\approx 3$ GeV/$c^2$ for the
mass lower limits.

\begin{table}[b]
\vspace{-3ex}
\caption{Theoretical cross sections ($\sigma_{\rm NLO}$ in fb~\cite{nloa,nlob}),
  expected $b'$ yield ($N_{exp}$) after selection, median expected 95\% C.L.
  limit ($\sigma_{\rm exp}$ in fb), and observed  95\% C.L.  limit ($\sigma_{\rm
    obs}$ in fb)  for \bprime at varying
  masses. $\sigma_{\rm NLO}$ and $N_{exp}$ have 10\% uncertainties.}
\begin{ruledtabular}
\begin{tabular}{lrrrrrrrrrrrr}
Mass [GeV/$c^2$]   & 260 & 300 & 325 & 350 & 375 & 400 & 425 \\
\hline
$\sigma_{\rm NLO}$  &630  & 227  &125&65&34&19&9.5\\
$N_{exp}$           &203.2& 91.5 & 53.0 & 28.1 & 14.9 & 7.6 & 4.0  \\
 $\sigma_{\rm exp}$   & 72 & 49 & 40 & 34 & 34 & 34 & 34  \\
    $\sigma_{\rm obs}$& 72 & 72 & 66 & 53 & 36 & 33
    & 34 \\ 
\end{tabular}
\label{table:bprres}
\end{ruledtabular}
\end{table}

In conclusion, we have searched for pair production of $b'$
quarks with subsequent decay to $tW$. Though there are events with
larger $H_T$ than expected in the 7-jet event distribution in Fig~\ref{fig:5j_njets}, we do
not see evidence of a signal. We calculate upper limits on the $b'$
pair production cross section ($\lesssim 30$ fb for $m_{b'}>$375
GeV/$c^2$) and set the most restrictive direct lower limit on the mass of a
down-type fourth-generation quark, increasing the limit by 34
GeV/$c^2$ to $m_{b'} \ge 372$ GeV/$c^2$ and significantly reducing the allowed mass range.  

We thank the Fermilab staff and the technical staffs of the participating institutions for their vital contributions. This work was supported by the U.S. Department of Energy and National Science Foundation; the Italian Istituto Nazionale di Fisica Nucleare; the Ministry of Education, Culture, Sports, Science and Technology of Japan; the Natural Sciences and Engineering Research Council of Canada; the National Science Council of the Republic of China; the Swiss National Science Foundation; the A.P. Sloan Foundation; the Bundesministerium f\"ur Bildung und Forschung, Germany; the World Class University Program, the National Research Foundation of Korea; the Science and Technology Facilities Council and the Royal Society, UK; the Institut National de Physique Nucleaire et Physique des Particules/CNRS; the Russian Foundation for Basic Research; the Ministerio de Ciencia e Innovaci\'{o}n, and Programa Consolider-Ingenio 2010, Spain; the Slovak R\&D Agency; and the Academy of Finland. 

\begin{thebibliography}{99}
\bibitem{chiral} We specify chiral quarks to distinguish from theories
  of a vector-like fourth generation.
\bibitem{coordinates}
CDF uses a cylindrical coordinate system with
  the $z$ axis along the proton beam axis. Pseudorapidity is $\eta
  \equiv -\ln(\tan (\theta/2))$, where $\theta$ is the polar angle
  relative to the proton beam direction, and $\phi$ is the azimuthal angle  while $p_T = |p|\ \sin\theta$, $E_T = E\ \sin\theta$.
\bibitem{metdef} 
Missing transverse momentum, $\missET$, is defined as the magnitude of the vector $ -\sum_i E_T^i \vec n_i$ where $E_T^i$ are the magnitudes of transverse energy contained in each calorimeter tower $i$, and $\vec n_i$ is the unit vector from the interaction vertex to the tower in the transverse ($x,y$) plane.
\end{thebibliography}

\end{document}